\documentclass[11pt]{IEEEtran}

\usepackage{amssymb}
\usepackage{color}
\usepackage{graphicx}
\usepackage{amsmath}
\usepackage{wasysym}
\usepackage{multirow}
\usepackage{booktabs}
\usepackage{lineno}
\usepackage{soul}
\usepackage[draft]{fixme}
\usepackage{longtable}
\usepackage{supertabular}
\usepackage[colon]{natbib}
\usepackage{authblk}

\def\qut#1{``#1''}

\newcommand{\delOx}{$\delta{^{18}\mathrm{O}}$ }
\newcommand{\delD}{$\delta\mathrm{D}$ }
\newcommand{\Dxs}{$\mathrm{D_{xs}}$ }

\begin{document}\sloppy

\title{Water isotopic ratios from a~continuously melted ice core sample}

\author[1]{V.~Gkinis%
\thanks{Correspondence to: V. Gkinis}%
\thanks{v.gkinis@nbi.ku.dk}}
\author[1]{T.~J.~Popp}
\author[1]{T.~Blunier}
\author[2]{M.~Bigler}
\author[2]{S.~Sch\"{u}pbach}
\author[1]{E.~Kettner}
\author[1,3]{S.~J.~Johnsen}

\affil[1]{Centre for Ice and Climate, Niels Bohr Institute, University of Copenhagen, Juliane Maries Vej 30, \newline 2100 Copenhagen, Denmark}
\affil[2]{Physics Institute, Climate and Environmental Physics and Oeschger Centre for Climate Change Research University of Bern, Sidlerstrasse 5, 3012 Bern, Switzerland}
\affil[3]{Science Institute, University of Iceland, Dunhaga 3, 107, Iceland}

%
%
%
%
%
%
%
%

\maketitle

\begin{abstract}
A~new technique for on-line high resolution isotopic analysis of
liquid water, tailored for ice core studies is presented.  We built an
interface between a  Wavelength Scanned Cavity Ring Down Spectrometer
(WS-CRDS) purchased from Picarro Inc.~and a~Continuous Flow Analysis~(CFA) system. The system
offers the possibility to perform simultaneuous water isotopic
analysis of $\delta^{18}$O and $\delta$D on a continuous stream of liquid water as
generated from a continuously melted ice rod. Injection of
sub ${\mu}$l amounts of liquid water is achieved by pumping
sample through a fused silica capillary and instantaneously vaporizing
it with 100\,\% efficiency in a~home made oven at a~temperature of
170\,$^\circ$C.  A calibration procedure allows for proper reporting
of the data on the VSMOW--SLAP scale.  We apply the necessary corrections
based on the assessed performance of the system regarding instrumental
drifts and dependance on the water concentration in the optical \mbox{cavity}.
The melt rates are
monitored in order to assign a depth scale to the measured isotopic
profiles.  Application of spectral methods yields the combined
uncertainty of the system at below 0.1\,{\permil} and 0.5\,{\permil}
for $\delta^{18}$O and $\delta$D, respectively. This performance is comparable to
that achieved with mass spectrometry. Dispersion of the sample in the
transfer lines limits the temporal resolution of the technique.  In this work
we investigate and assess these dispersion effects.  By using an
optimal filtering method we show how the measured profiles can be
corrected for the smoothing effects resulting from the sample
dispersion.  Considering the significant advantages the technique
offers, i.e.\ \mbox{simultaneuous} measurement of $\delta^{18}$O and $\delta$D,
potentially in combination with chemical components that are
traditionally measured on CFA systems, notable reduction on analysis
time and power consumption, we consider it as an alternative to
traditional isotope ratio mass spectrometry with the possibility to be
deployed for field ice core studies.  We present data acquired in the field
during the 2010 season as part of the NEEM deep ice core drilling project
in North Greenland.
\end{abstract}

\section{introduction}

Polar ice core records provide some of the most detailed views of past
environmental changes up to 800\,000\,yr before present, in large
part via proxy data such as the water isotopic composition and
embedded chemical impurities.  One of the most important features of
ice cores as climate archives, is their continuity and the potential
for high temporal resolution.  Greenland ice cores are particularly
well suited for high resolution paleoclimatic studies, because
relatively high snow accumulation rates allow seasonal changes in
proxy data to be identified more than 50\,000\,yr in the past
\citep{Johnsen1992, NGRIPmembers2004}.

The isotopic signature of polar precipitation, commonly expressed
through the $\delta$ notation\footnote{Isotopic abundances are
typically reported as deviations of a~sample's isotopic ratio relative
to that of a~reference water (e.g.\ VSMOW) expressed in per mil (\permil) through the
$\delta$ notation: $\delta^{i} = \left(
\displaystyle{\frac{^{i}{R}_{\textrm{sample}}}
{^{i}{R}_\textrm{{SMOW}}}} - 1 \right) \times 1000 $ where
$^{2}{R} =
\displaystyle{\frac{^{2}\textrm{H}}{^{1}\textrm{H}}}$ and
$^{18}{R} =
\displaystyle{\frac{{^{18}O}}{{^{16}O}}}$}
\citep{Epstein1953, Mook2000} is related to the temperature gradient
between the evaporation and condensation site \citep{Dansgaard1964}
and has so far been used as a~proxy for the temperature of the cloud
at the time of condensation \citep{Jouzel1984, Jouzel1997,
Johnsen2001}. One step further, the combined signal of {\delD} and
{\delOx} commonly referred to as the deuterium excess (hereafter
{\Dxs}), constitutes a~useful paleothermometer tool. Via its high
correlation with the temperature of the evaporation
source \citep{Johnsen1989}, it has been used to resolve issues related
to changes in the location of the evaporation site \citep{Cuffey2001,
Kavanaugh2002}. A~relatively recent advance in the use of water
isotope ratios as a~direct proxy of firn temperatures, has been introduced
by \cite{Johnsen2000}.  Assessment of the diffusivity of the water
isotopologues in the porous medium of the firn column can yield
a~temperature history, provided a~dating model is available.

The measurement of water stable isotopic composition is typically
performed off-line via discrete sampling with traditional isotope
ratio mass spectrometry (hereafter IRMS).  While high precision and
accuracy can routinely be achieved with IRMS systems, water isotope
analysis remains an elaborate process, which is demanding in terms of
sample preparation, power consumption, sample size, consumables, isotope
standards and carrier gases.  The analysis of a~deep ice core at its
full length in high resolution (typically 2.5 to 5\,cm per sample)
requires the processing of a~vast amount of water samples and can take
years to complete.  Additionally, these procedures often come at the
expense of not fully exploiting the temporal resolution available in
the ice core.

Laser spectroscopy in the near and mid infrared region has been
demonstrated as a~potential alternative for water isotope analysis,
presenting numerous advantages over IRMS \citep{Kerstel1999,
Kerstel2004}.  A~major advantage of the technique is the ability to
directly inject the sampled water vapour in the optical cavity of the
spectrometer where both isotopic ratios $^{18}\mathrm{O}/^{16}\mathrm{O}$
and $^{2}\mathrm{H}/^{1}\mathrm{H}$ are measured simultaneuously. In
contrast, in the most common IRMS techniques water is not measured as
such, but has to be converted to a~different gas prior to measurement.
For {\delOx} analysis, the $\mathrm{CO}_2$ equilibration method
\citep{Epstein1953} has been widely used, whereas {\delD} analysis
commonly involves the reduction of water to hydrogen gas over hot
uranium \citep{Bigeleisen1952, Vaughn1998, Huber2003}, or chromium
  \citep{Gehre1996}. However, the
combined use of these two methods rules out simultaneous analysis of
both water isotopologues on a~given sample.  More recently, in
combination with the use of continuous flow mass spectrometers,
conversion of water to $\text{CO}$ and $\text{H}_2$ is performed in a~pyrolysis
furnice \citep{Begley1997} and allows simultaneous {\delD} and {\delOx}
measurement, but still on a~single discrete sample. One of the drawbacks
  of this technique is the interference of $\mathrm{NO}$ , formed at the ion
  source by the reaction of $\text{N}_2$ and $\text{O}_2$ with the $\text{CO}$
  signal at {\it m/z}\,$=$\,30 \citep{Accoe2008}.
  Nowadays,
commercial IR spectrometers are available with a~precision comparable
to IRMS systems \citep{Lis2008, Brand2009}. These units typically
receive a~continuous stream of water vapor and offer ease of use and
portability.

The analysis of another set of ice core proxies, that of chemical
impurities, has similarly been an elaborate process, traditionally
performed with liquid \mbox{chromatography} techniques.  With the advent of
Continuous Flow Analysis (heareafter CFA) from continuously melted ice
core segments, the measurement of chemical impurities has reached the
point of largely exploiting the high resolution available in the core
while it is often performed in the field \citep{Sigg1994,
Rothlisberger2000, Kaufmann2008}. The continuous, on-line nature of
the technique has resulted in a~considerable reduction in sample
preparation and processing times.  Recently,
\cite{Schupbach2009} demonstrated the measurement of $\mathrm{CH}_4$ mixing
ratios in an on-line semi continuous mode with the use of a~gas
chromatograph combined with a~pulsed discharge and a~thermal
conductivity detector.

Here, we demonstrate the ability to perform continuous measurements of
water isotope ratios from a~stream of water vapor derived from
a~continuously melting ice rod by coupling a~commercial IR
spectrometer to a~CFA system via a~passive, low volume flash
evaporation module.  In the following, we assess the system's
precision, accuracy, and efficient calibration.  We then comment on
issues related to sample dispersion in the sample transfer lines, the
evaporation module and the optical cavity of the spectrometer itself
in order to determine the expected smoothing imposed on the acquired
data sets. Finally, isotopic analysis of ice core samples from the
NEEM deep ice core are presented and compared to measurements
performed in discrete mode.

\section{Experimental}

\subsection{Continuous flow analysis}

In the system described here, (Fig.~\ref{Fig1}) an ice rod measuring
3.2\,$\times$\,3.2\,$\times$\,110\,cm (hereafter CFA run) is
continuously melted on a~copper, gold-nickel coated melter at
a~regulated temperature of 20\,$^\circ$C.  The concentric arrangement
of the melter's surface facilitates the separation of the sample that
originates from the outer and inner part of the core.  Approximately
90\,{\%} of the sample from the inner part is transfered to the
analytical system by means of a~peristaltic pump with a~flow rate of
16\,ml\,min$^{-1}$. This configuration provides an overflow of $\approx$10\,{\%} from the inner to the outer part of the melter and ensures
that the water sample that is introduced into the analytical system is
not contaminated.

A~stainless steel weight sitting on top of the ice rod enhances the
stability and continuity of the melting process. An optical encoder
connected to the stainless steel weight, records the displacement of
the rod. This information is used to accurately define the depth scale
of the produced water isotope data.  Breaks in the ice rod are logged
prior to the melting process and accounted for, during the data
analysis procedure.

Gases included in the water stream originating from the air bubbles in
the ice core are extracted in a~sealed debubbler, with a~volume of
$\approx$300\,$\mu$l. The melt rate of the present system is
approximately 3.2\,cm\,min$^{-1}$, thus resulting in an analysis time
of $\approx$35\,min per CFA run.  During the intervals between CFA
runs, mQ\footnote{filtered and deionized water with a resistivity more than
  $18.2 ~\mathrm{M\Omega\,\,\, cm}$ and a total organic content less than 10\,ppb.}
 water is pumped through the system. A~4-port injection valve
(V1 in Fig.~\ref{Fig1}) allows the selection between the mQ and sample
water. The mQ water is spiked with isotopically enriched water
containing 99.8\,atom\,\% deuterium, Cortecnet Inc.) in a~mixing ratio of
$\approx $1\,ppm. In this way a~distinction between sample and mQ
water is possible, facilitating the identification of the beginning
and end times of a~CFA run.

For further details on the analysis of chemical components or the
extraction of gases for greenhouse gas measurements the reader is
refered to \cite{Kaufmann2008} and \cite{Schupbach2009}.

\subsection{The water isotope measurement}

We follow the same approach as previously presented in
\cite{Gkinis2010} by coupling a~commercially available Cavity Ring
Down IR spectrometer (hereafter WS-CRDS) purchasecd from Picarro Inc.
(Picarro L1102-i) \citep{Crosson2008}.  The spectrometer operates with
a~gas flow rate of 30\,standard\,ml\,min$^{-1}$. In the optical cavity the
pressure is regulated at 47\,mbar with two proportional valves in
a~feedback loop configuration up- and down-stream of the optical
cavity at a~temperature of 80\,$^\circ$C.  The high signal to noise
ratio achieved with the Cavity Ring Down configuration in combination
with fine control of the environmental parameters of the spectrometer,
result in a~performance comperable to modern mass spectrometry systems
taylored for water stable isotope analysis.

\begin{figure*}
\vspace*{2mm}
\center
\includegraphics[width=150mm]{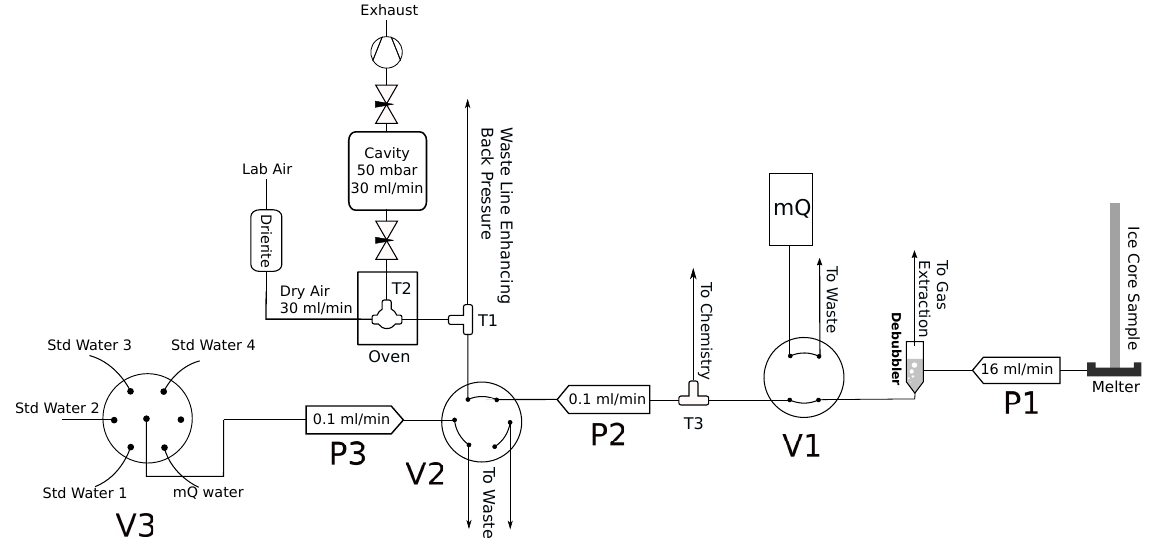}
\caption{Block diagram of the CFA-CRDS system.}
\label{Fig1}
\end{figure*}

A~6-port injection valve (V2 in Fig.~\ref{Fig1}) selects sample from
the CFA line or a~set of local water standards. The isotopic
composition of the local water standards is determined with
conventional IRMS and reported with respect to VSMOW standard. A~6-port
selection valve (V3 in Fig.~\ref{Fig1}) is used for the switch between
different water standards.  A~peristaltic pump (P3 in Fig.~\ref{Fig1})
in this line with variable speeds, allows adjustment of the water
vapor concentration in the spectrometer's optical cavity, by varying
the pump speed. In that way, the system's sensitivity to levels of different
water concentration can be investigated and a~calibration procedure can be
implemented.  We use high purity Perfluoroalkoxy~(PFA) tubing for all
sample transfer lines.

Injection of water sample into the evaporation oven takes place via
a~{$\varnothing$} 40\,$\mu$m fused silica capillary where
immediate and 100\,{\%} evaporation takes place avoiding any
fractionation effects. The setpoint of the evaporation temperature is
set to 170\,$^\circ$C and is regulated with a~PID controller. The
amount of the injected water to the oven can be adjusted by the
pressure gradient maintained between the inlet and waste ports of the
T1 tee-split (Fig.~\ref{Fig1}). The latter depends on the ratio of the
inner diameters of the tubes connected to the two ports as well as the
length of the waste line. The total water sample consumption is
$\approx$0.1\,ml\,min$^{-1}$ maintained by the peristaltic pump P2
(Fig.~1).  For a~detailed description of the sample preparation and
evaporation module the reader may reffer to \cite{Gkinis2010}.
A~smooth and undisturbed sample delivery to the spectrometer at the
level of $\approx$20\,000\,ppm results in optimum performance of the
system. Fluctuations of the sample flow caused by air bubbles or
impurities are likely to result in a~deteriorated performance of the
measurement and are occasionally observed as extreme outliers on both
{\delOx} and {\delD} measurements. The processes that control the
occurence of these events are still not well understood.

\section{Results and discussion -- from raw data to isotope records}

In this study we present data collected in the framework of the NEEM
ice core drilling project. Measurements were carried out in the field
during the 2010 field season and span 919.05\,m of ice core (depth
interval 1281.5--2200.55\,m). Here we exemplify the performance of the
system over a~section of ice from the depth interval 1382.152--1398.607.
  The age of this section spans
$\approx$411\,yr with a~mean age of 10.9\,ka\,b2k
  \footnote{thousand years before 2000~AD (ka b2k)}. The reported age is
based on a~preliminary time scale constructed by stratigraphic
transfer of the GICC05 time scale \citep{Rasmussen2006} from the NGRIP
to the NEEM ice core.

\begin{figure*}
\vspace*{2mm}
\center
\includegraphics[width=150mm]{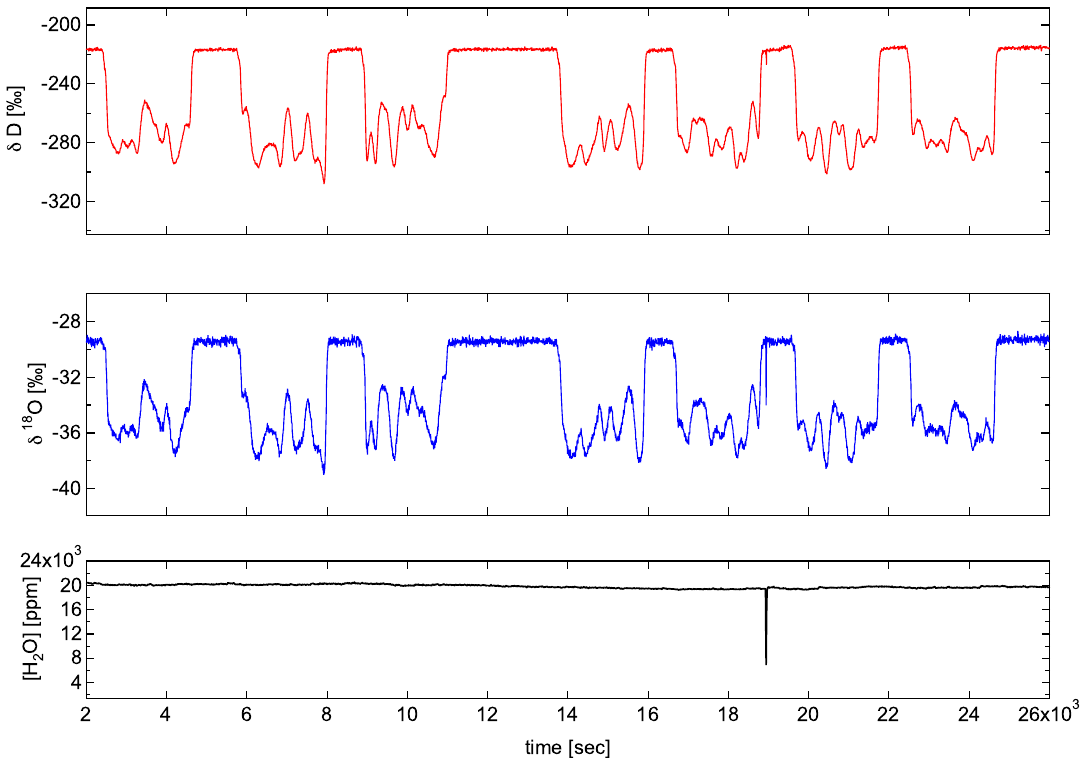}
\caption{Raw signals spanning 7 CFA runs on 29~May 2010.}
\label{Fig2}
\end{figure*}

In Fig.~\ref{Fig2} we present an example of raw data as acquired by
the system. This data set covers 7~CFA runs (7.70\,m of ice).  A~clear
baseline of the isotopically heavier mQ water can be seen in between
CFA runs.  At $t=1.9 \times 10{^4}$\,s one can observe a~sudden drop
in the signal of the water concentration due to a~scheduled change of
the mQ water tank. Adjacent to this, both {\delOx} and {\delD} signals
present a~clear spike, characteristic of the sensitivity of the system
to the stability of the sample flow rates.

\subsection{VSMOW -- water concentration calibrations}

Before any further processing we correct the acquired data for
fluctuations of the water concentration in the optical \mbox{cavity}.  To
a~good approximation the \delOx~ and \delD~ signals show a~linear response to differences
in water concentrations around 20\,000\,ppmv \citep{Brand2009,
Gkinis2010}. A~correction is performed as:
\begin{equation}
\label{Eq.4}
\Delta \delta = \alpha (R_{20} - 1) {}
\end{equation}
Here $R_{20} = \frac{[\mathrm{H}_{2}\mathrm{O}]}{20\,000}$, $\alpha_{18} =
1.94$\,{\permil} and $\alpha_{\rm D} = 3.77$\,{\permil} as estimated in
\cite{Gkinis2010}.
The estimation of these values has been performed several times during the period
July~2009--October~2011. Based on compiled data from 6~calibrations
we can report that these values do not appear to be drifting in the course of the
two years. The values reported here are the ones estimated chronologically closer
to the measurements we present here.
These values are most likely instrument specific and should be used with caution
for other analyzers.
We typically operate the system in the area of 17\,000--22\,000\,ppmv in which
we observe no impact of the water concentration level on the precision of the
isotopic signal. The mean and standard deviation of the water concentration signal
in the course of aprroximately 7\,h as seen in Fig.~2 is $19\;939 \pm 306$\,ppmv.
The
water concentration correction is applied to the raw data, scaling all
the isotopic values to the level of 20\,000\,ppmv.

Raw data are expressed in per mil values for both {\delOx} and {\delD} and
ppmv for the water vapour concentration. These values are based on the
slope and intercept values of the instrument's stored internal
calibration line. Due to apparent instrumental drifts though, the
latter are expected to deviate with time. To overcome this problem we
perform frequent VSMOW calibrations by using 3~local water standards
with well known {\delOx} and {\delD} values measured by conventional
Isotope Ratio Mass Spectrometry combined with a~pyrolysis glassy
carbon reactor (Thermo DeltaV--TC/EA). The water standards are
transported and stored in the field using the necessary precautions
to avoid evaporation. We used amber glass bottles with silicon sealed
caps. Two of the water standards
are used to calculate the slope and the intercept of the calibration
line and the third is used for a check of the linearity and the
accuracy. The frequency of the VSMOW calibrations depends on the
work flow of the overall CFA system. For the particular section we
present here a VSMOW calibration was performed the same day.

\begin{figure*}
\vspace*{2mm}
\center
\includegraphics[width=160mm]{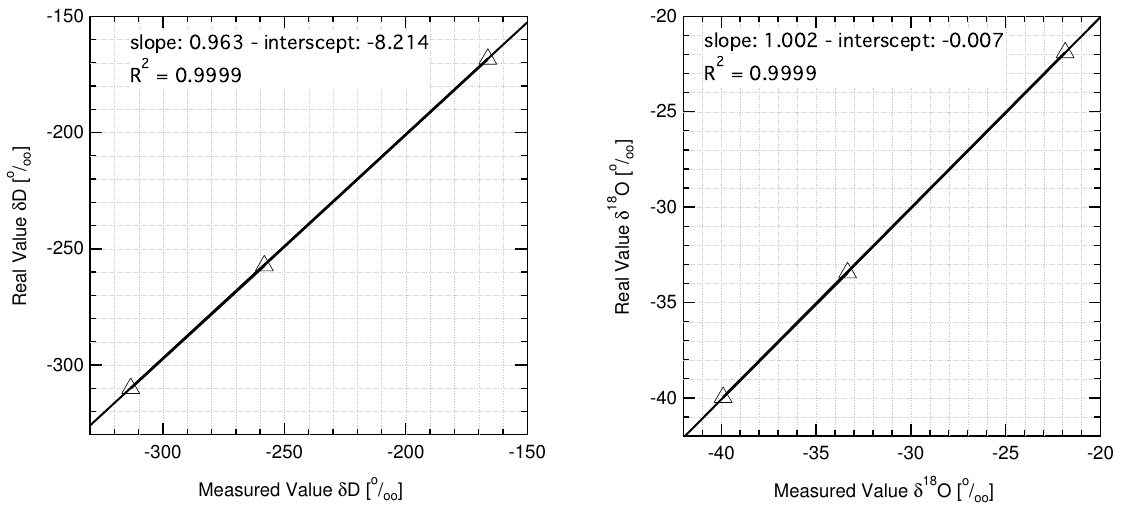}
\caption{VSMOW calibrations with three water standards ``$-$22'', ``$-$40''
      and ``NEEM''; Left -- \delD , right -- \delOx.}\vspace{4mm}
\label{Fig3}
\end{figure*}

In Fig.~\ref{Fig3} we illustrate this calibration. Based on the
measured and real values of the standards ``$-$22'' and ``$-$40'' the
measurement of the ``NEEM'' standard can be used as a test for
accuracy and linearity. In Table~\ref{table01} we present the
VSMOW calibrated values of the water standards as estimated with
the IRMS  system. The slope and intercept of the calibration lines
are [1.002, $-$0.007] for \delOx ~and [0.963, $-$8.214] for \delD. Based
on these values one can calculate the isotopic composition of the ``NEEM''
standard. The results we obtain are $-33$.4\,{\permil} and $-256$.99\,{\permil}
for \delOx ~and \delD~ respectively. This results in a difference
of $0.04$\,{\permil} ($0.$31\,{\permil}) for \delOx~ (\delD), giving an indication
about the accuracy obtained by the system.

\begin{table}[t]
\caption{Isotopic composition of the standards used for the VSMOW calibrations in {\permil}.}
\vskip4mm
\centering
\begin{tabular}{rrrr}
\hline
 & $-$22 & NEEM & $-$40\\
\hline
\delOx & $-$21.9 & $-$33.44 & $-$39.97\\
\delD & $-$168.4 & $-$257.3 & $-$310\\
\hline
\end{tabular}
\label{table01}
\end{table}

\subsection{The depth scale}

\begin{figure*}
\vspace*{2mm}
\center
\includegraphics[width=120mm]{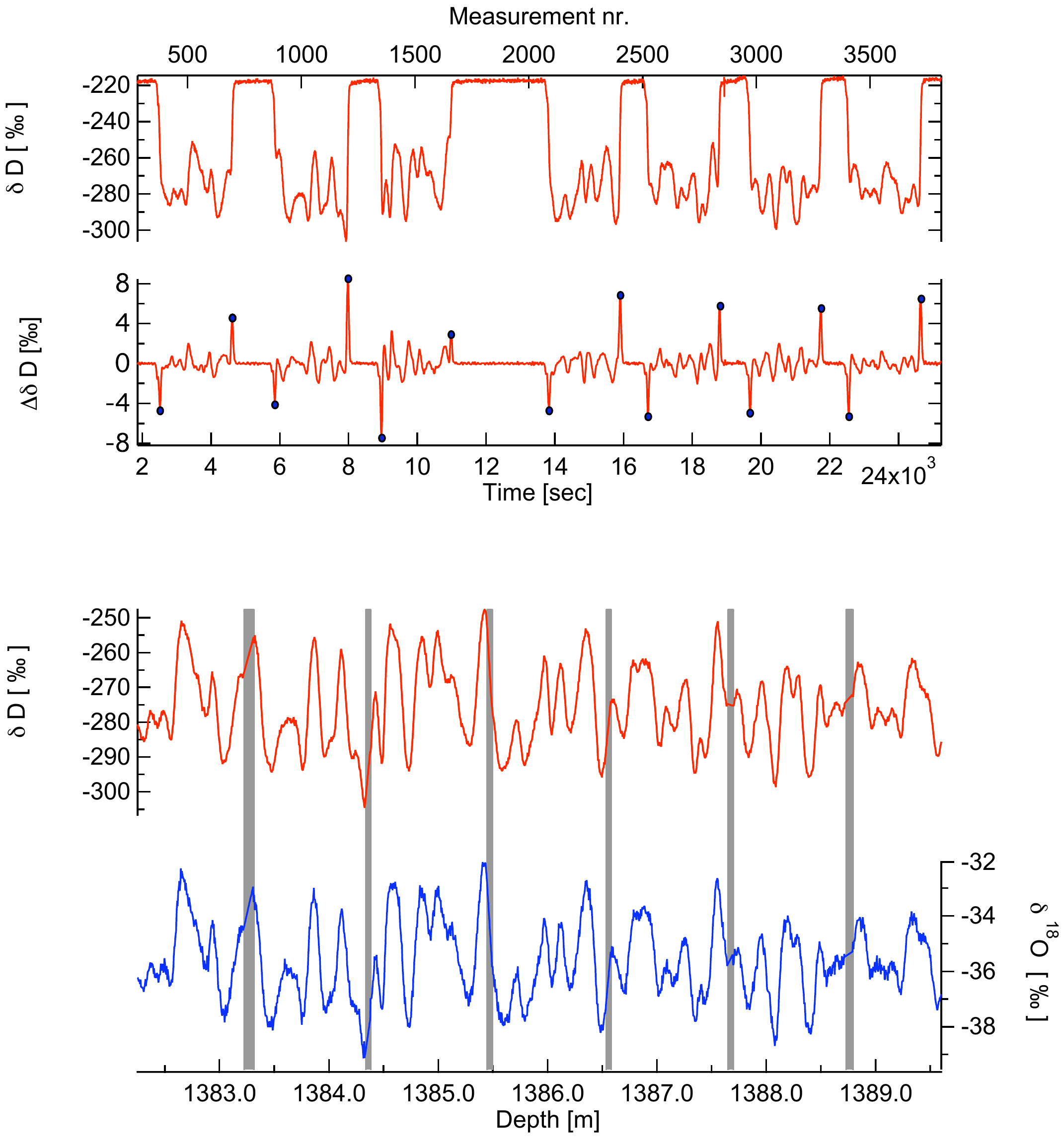}
\caption{The beginning and end of each CFA run is determined by the extrema of the 1{st} derivative of
the isotopic signal, presented on the top graph. With gray bars we indicate the position and width of sections with
data that are missing due to breaks in the ice, or removed in order to account for the transition from mQ water to sample and vice
cersa.}
\label{Fig4}
\end{figure*}

The melting process is recorded by an optical encoder connected to the
top of the stainless steel weight that lies on top of the ice rod. The
data acquired by the optical encoder allow for a~conversion of the measurement
time scale to a~depth scale. In order to locate the beginning and end
of every run we take advantage of the isotopic step observed during
the transition between mQ baseline and sample water.  A~smoothed
version of the discrete derivative of the acquired isotope data for
both {\delOx} and {\delD} reveals a~local minimum (maximum) for the
beginning (end) of the measurement (Fig.~\ref{Fig4}). The logged depth
of the top and the bottom of the CFA run is assigned to these points.
Data that lie in the transition interval between mQ and sample water
are manually removed from the series.  Additional breaks within a~CFA
run that can possibly be created during the drilling or processing
phase of the ice core, are taken into account at the last stage of the
data analysis. If necessary and depending on their size, the gaps can
be filled by means of some interpolation technique. Here, due to the
small size of the gaps we use a~linear interpolation scheme. The use of
more advanced methods is also possible but is out of the scope of this
work.  The processed profiles presented in Fig.~\ref{Fig4} are
reported with a~nominal resolution of 5\,mm. The interpolated sections
are highlighted with gray bars.  Their width indicates the length of
the gaps.

\subsection{Noise level -- comparison with discrete data}

\begin{figure*}
\vspace*{2mm}
\center
\includegraphics[width=120mm]{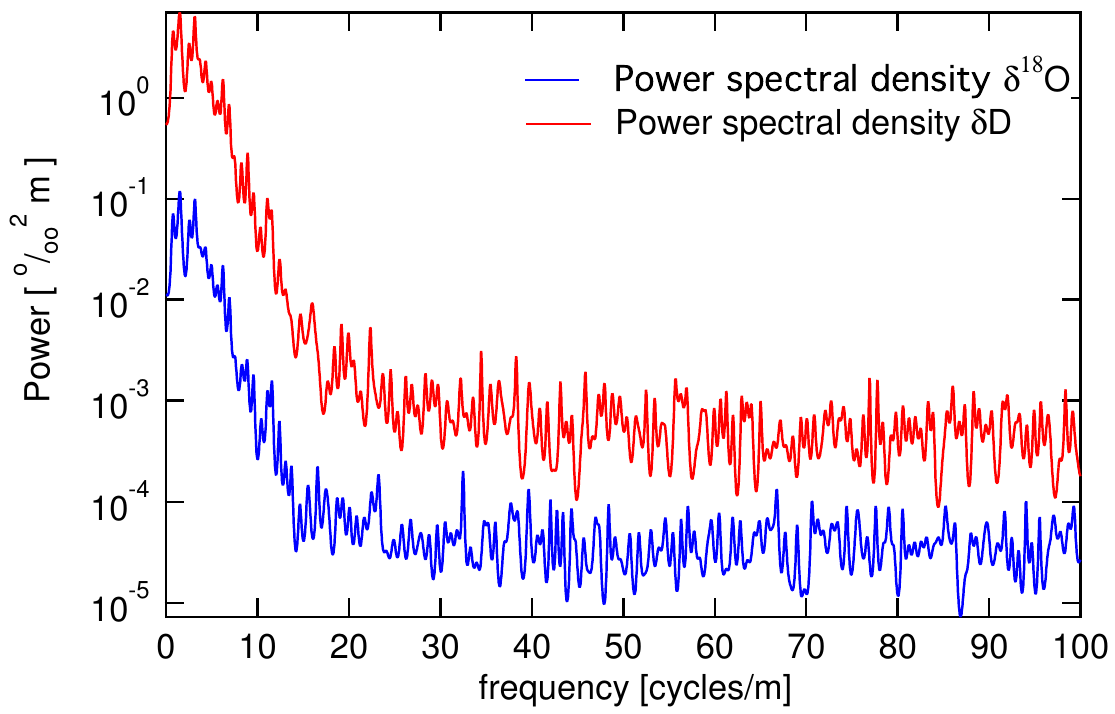}
\caption{{\delOx} and {\delD} power spectral density based on the data
of the interval 1382.152--1398.607\,m.}
\label{Fig5}
\end{figure*}

An estimate of the noise level of the measurements can be obtained
from the appropriately normalized power spectral density of the time
series.  Using the \delOx ~and \delD ~data of the section under consideration,
  we implement an autoregressive spectral estimation
method developed by \cite{Burg1975} by the use of the algorithm
introduced by \cite{Andersen1974}.  The order of the autoregressive
model is $M$\,$=$\,300.  The standard
deviation of the time series will be defined as:
\begin{equation}
\sigma^2 = \int_{-f_{c}}^{f_{c}} | \hat{\eta} (f) |^2 df {}
\end{equation}
where the Nyquist frequency is $f_c = 100$\,cycles\,m$^{-1}$ and $|
\hat{\eta} (f) |^2$ can be obtained by a~linear fit on the flat high
frequency part of the spectrum (Fig.~\ref{Fig5}).  By performing this
analysis we obtain $\sigma_{18} = 0.055$\,{\permil} and $\sigma_{\rm D} =
0.21$\,{\permil}.

\begin{figure*}
\vspace*{2mm}
\center
\includegraphics[width=120mm]{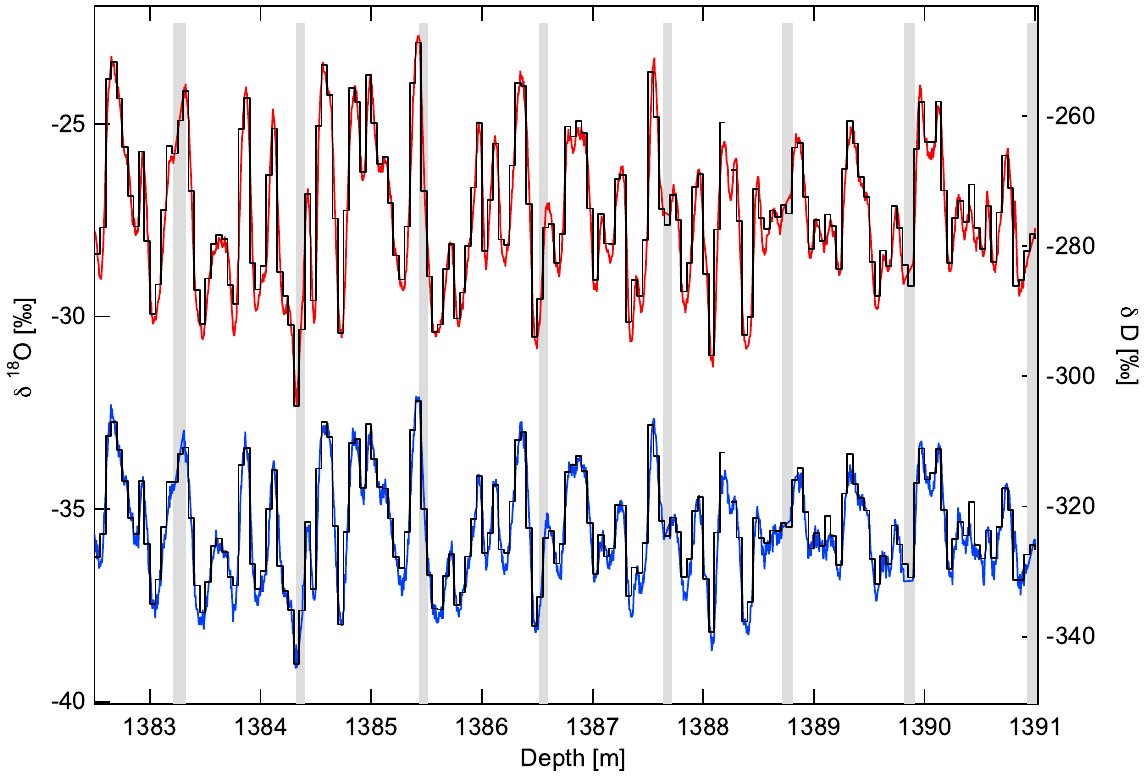}
\caption{Comparison CRDS-CFA with 5\,cm discrete samples for {\delD} (top) and {\delOx} (bottom).
 Bars indicate the position and width of sections with missing/removed data.}
\label{Fig6}
\end{figure*}

In order to validate the quality of the calibrations as well as the
estimated depth scale we compare the CFA data with measurements
performed in a~discrete fashion using the same WS-CRDS spectrometer in
combination with a~sample preparation evaporator system
\citep{Gupta2009} and an autosampler.  The discrete samples are cut in
a~resolution of 5\,cm. The sample injection sequence takes into
account apparent memory effects and results are reported on the VSMOW
scale by appropriate calibration using local water standards.  The
results are illustrated in Fig.~\ref{Fig6} for {\delOx} and {\delD}. The
comparison of the data sets demonstrates  validity of the followed calibration
procedures. The benefits of the technique in terms of achieved
resolution can be seen when one compares the two datasets over
isotopic cycles with relatively small amplitude and higher
frequency. Such an example can be seen at the depth of 1390.5\,m
where a~sequence of 4~cycles is sampled relatively poorly with the
discrete method when compared to the on-line system.  This performance
can benefit studies that look into the spectral properties of the
signals by providing better statistics for the obtained measurements.

\subsection{Obtained resolution -- diffusive sample mixing}\label{section_resolution}

One of the advantages of the combined CFA--CRDS technique for water
isotopic analysis of ice cores lies in the potential for higher
resolution measurements relative to discrete sampling.  However,
diffusion effects in both the liquid and the vapor phase are expected
to attenuate the obtained resolution.

Attenuation of the initial signal of the precipitation occurs also via
a~combination of in situ processes that take place after deposition.
The porous medium of the firn column allows for an exchange of water
molecules in the gas phase along the isotopic gradients of the
profile. For the case of polar sites, this process has been studied
extensively \citep{Johnsen1977, Whillans1985} and can be well
described and quantified provided that a~good estimate of the
diffusivity coefficient and a~strain rate history of the ice core site
are available \citep{Johnsen2000}. The process ceases when the porous
medium is closed-off and the diffusivity of air reaches zero at
a~density of $\approx$804\,kg\,m$^{-3}$.  Deeper in the ice,
diffusion within the ice crystals takes place via a~process that is
considerably slower when compared with the firn diffusion.  At
a~temperature of $-$30\,$^\circ$C the diffusivity coefficients of
these two processes differ by 4~orders of magnitude
\citep{Johnsen2000}.

Assuming an isotopic signal ${\delta}_{\rm pr}$ for the precipitation,
the total effect of the diffusive processes, in-situ and experimental,
can be seen as the convolution of ${\delta}_{\rm pr}(z)$ with
a~smoothing filter ${\mathcal{G}}_{\rm tot}$.
\begin{equation}
\label{Eq.2}
\delta_{\rm m}(z) = \int_{-\infty}^{\infty}{\delta}_{\rm pr}(\tau ) {\mathcal{G}}_{\rm tot} (z - \tau ) d \tau =
[ \delta_{\rm pr} \ast {\mathcal{G}}_{\rm tot} ] (z) {}
\end{equation}
where $\delta_{\rm m} (z)$ is the measured signal and ($\ast$) denotes the
convolution operation.  Since instrumental and
in-situ firn-ice diffusion are statistically independent, the variance
of the total smoothing filter is the sum of the variances of the
in-situ and experimental smoothing filters (hereafter
$\mathcal{G}_{\rm firn}$, $\sigma_{\rm firn}$, $\mathcal{G}_{\rm
cfa}$, $\sigma_{\rm cfa}$).
\begin{equation}
\label{filter_variance}
\sigma_{\rm tot}^{2} = \sigma_{\rm firn}^2 + \sigma_{\rm cfa}^2 {}
\end{equation}
It can be seen that any attempt to study firn and ice diffusion by
means of ice core data obtained with an on-line method similar to the
one we present here, requires a~good assesment of the diffusive
properties of the experimental system. The latter is possible if one
is able to estimate the variance of the smoothing filter
$\mathcal{G}_{\rm cfa}$ expressed by the variance $\sigma_{\rm cfa}^2$
(hereafter diffusion length).

One way to approach this problem is to measure the response of the
system to a~step function. Ideally, in the case of zero diffusion,
a~switch between two isotopic levels would be described by a~scaled
and shifted version of the the Heaviside unit step function as:
\begin{equation}\label{Eq.3}
\delta_{\rm H}(z) = \left\{
\begin{array}{ll}
C_{2}& \qquad z<0\\ C_{1} H (z) + C_{2}& \qquad z\geq 0
\end{array} \right. {}
\end{equation}
where the isotopic shift takes place at $z = 0$, $H(z)$ is the
Heaviside unit step function and $C_{1}$ and $C_{2}$ refer to the
amplitude and base line level of the isotopic step. Convolution of the
signal of Eq.~(\ref{Eq.3}) with $\mathcal{G}_{\rm cfa}$ and subsequent
calculation of the derivative yields,
\begin{equation}
\label{Eq.4}
\frac{d\delta_{\rm m}}{dz} = \frac{d\delta_{\rm H}}{dz} \ast \mathcal{G}_{\rm cfa} =
C_{1}\frac{dH}{dt} \ast \mathcal{G}_{\rm cfa} =
C_{1}\delta_{\rm Dirac} \ast \mathcal{G}_{\rm cfa} {}
\end{equation}
Thus the derivative of the measured signal, properly normalized,
equals the impulse respone of the system.  Applying the Fourier
transform, denoted by the overhead hat symbol, in Eq.~(\ref{Eq.4}),
and by using the convolution theorem, we deduce the transfer function
$\hat{\mathcal{G}}_{\rm cfa}$ of the system:
\begin{equation}
\widehat{\frac{d\delta_{\rm m}}{dz}} = C_{1} \hat \delta_{\rm Dirac}
\cdot \hat{\mathcal{G}}_{\rm cfa} = C_{1}\cdot \hat{\mathcal{G}}_{\rm cfa} {}
\end{equation}

\begin{figure*}
\vspace*{2mm}
\center
\includegraphics[width=150mm]{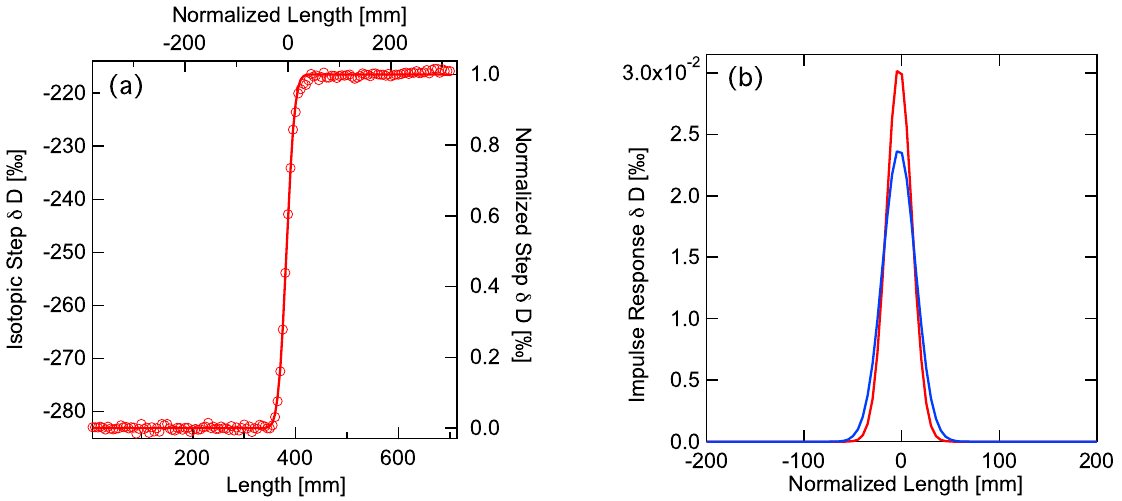}
\caption{\textbf{(a)}~Isotopic {\delD} step. The length scale is normalized so normalized
length\,$=$\,0 when the normalized {\delD} value equals 0.5\,{\permil}. \textbf{(b)}~Impulse
Response of the system for {\delD} based on the step response (red) and the spectral analysis
(blue) with $\sigma_{\rm cfa} = 13.4$ and 16.4\,mm, respectively.}
\label{Fig7}
\end{figure*}

In the case of the system presented here, an isotopic transition can
be observed when the main CFA valve (V1 in Fig.~\ref{Fig1}) switches
between mQ water and sample at the beginning and the end of each CFA
run as shown in Fig.~\ref{Fig4}. By using these transitions we are
able to construct isotopic steps and estimate the impulse response of
the system. Such an isotopic step is illustrated in Fig.~\ref{Fig7}a.
We fit the data of Fig.~\ref{Fig7}a with a~scaled version of the
cumulative distribution function of a~normal distribution described as
\begin{equation}
\delta_{\rm model} (z) = \frac{C_1^{\prime}}{2}
\left[
1 + {\rm erf} \left( \frac{z - z_0}{\sigma_{\rm step} \sqrt{2}} \right)
\right] + C_2^{\prime} {}
\end{equation}
The values of $C_1^{\prime}$, $C_2^{\prime}$, $z_0$ and $\sigma_{\rm
step}$ are estimated by means of a~least square optimization and used
accordingly to normalize the length scale and the isotopic values of
the step. A~nominal melt rate of 3.2\,cm\,min$^{-1}$ is used for all
the calculations presented here. We focus our analysis on the {\delD}
signal. The same approach can be followed for {\delOx}.  In
Fig.~\ref{Fig7}b we present the calculated impulse response of the
system.  The latter can be well approximated by a~Gaussian type filter
described as:
\begin{equation}
\label{Gaussian_filter}
\mathcal{G}_{\rm cfa} (z) = \frac{1}{\sigma_{\rm cfa}
\sqrt{2\pi}}{e}^{-\frac{z^2}{2\sigma_{\rm cfa}^{2}}} {}
\end{equation}
The diffusion length term $\sigma_{\rm cfa}$ is equal to $13.4 \pm
0.17$\,mm ($1\sigma$) as calculated with the least squares
optimization.  The transfer function for this filter will be given by
its Fourier transform, which is itself a~Gaussian and is equal to
\citep{Abramowitz1964}:
\begin{align}
\label{ftrans_gaussian}
&\mathfrak{F}
[ \mathcal{G}_{\rm cfa} (z) ] =
\hat{\mathcal{G}}_{\rm cfa} \nonumber \\ &=
\frac{1}{2\pi}\int_{-\infty}^{\infty} \frac{1}{\sigma_{\rm cfa}
\sqrt{2\pi}} {e}^{-\frac{z^2}{2\sigma_{\rm cfa}^2}} {e}^{- i k z}dk =
{e}^{\frac{-k^2 \sigma_{\rm cfa}^2}{2}}
\end{align}
where $k = \frac{2 \pi}{\lambda}$ and $\lambda$ is the wavelength\footnote{Here
the term wavelength refers to the isotopic signal in the ice and
should not be confused with the wavelength of the light emitted by the laser diode
of the spectrometer.} of a harmonic of the isotopic
signal. Harmonics with an initial amplitude $A_0$ and
wavenumber $k$ will be attenuated to a~final amplitude equal to:
\begin{equation}
\label{attenuation}
A = A_0 {e}^{\frac{-k^2 \sigma_{\rm cfa}^2}{2}} {}
\end{equation}

\begin{figure*}
\vspace*{2mm}
\center
\includegraphics[width=150mm]{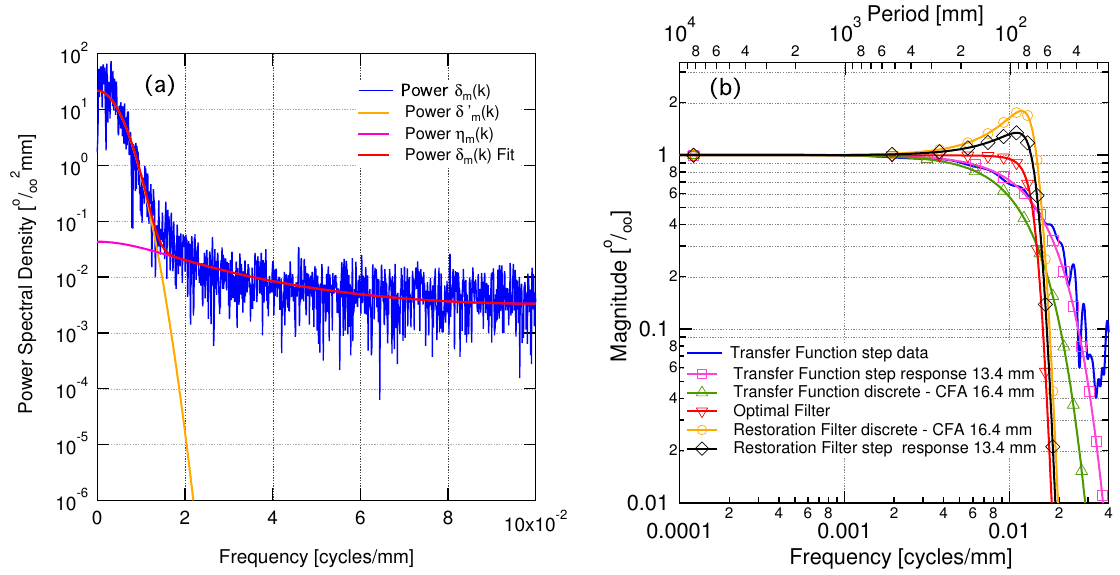}
\caption{\textbf{(a)}~Power spectral density of {\delD}. \textbf{(b)}~Transfer function calculated based on
the step response with $\sigma_{\rm cfa} = 13.4$\,mm (pink -- squares) and the comparison between
 discrete and CFA analysis with $\sigma_{\rm cfa} = 16.4$\,mm (green -- triangles). Restoration
 filters built considering the two different transfer functions are illustrated with orange
  circles ($\sigma_{\rm cfa} = 16.4$\,mm) and black diamonds ($\sigma_{\rm cfa} = 13.4$\,mm).}
\label{Fig8}
\end{figure*}

An estimate of the transfer function based on the data and the
cumulative distribution model is presented in Fig.~\ref{Fig8} (blue
and pink curve, respectively).  As seen in this plot, cycles with
wavelengths longer than 25\,cm experience negligible attenuation,
whereas cycles with a~wavelength of 7\,cm are attenuated by $\approx
50$\,{\%}.

The step response approach has been followed in the past for on-line
chemistry data.  In some studies such as \cite{Sigg1994} and
\cite{Rasmussen2005}, the resolution of the experimental system was
assessed via the estimation of the transfer function.  In other
studies \citep{Rothlisberger2000, Kaufmann2008}, the characteristic
time in which a~step reaches a certain level (typically $1/e$) with
respect to its final value, is used as a~measure of the obtained
resolution of the system.  A common weakness of this approach as
applied in the current, as well as previous studies, is that it is
based on the analysis of a~step that is introduced in the analytical
system by switching a~valve that is typically situated downstream of
the melting and the debubbling system. Consequently, the impact of
these last two elements on the smoothing of the obtained signals is
neglected.  In this study, this is the valve V1 in Fig.~\ref{Fig1}.

\begin{figure*}
\vspace*{2mm}
\center
\includegraphics[width=120mm]{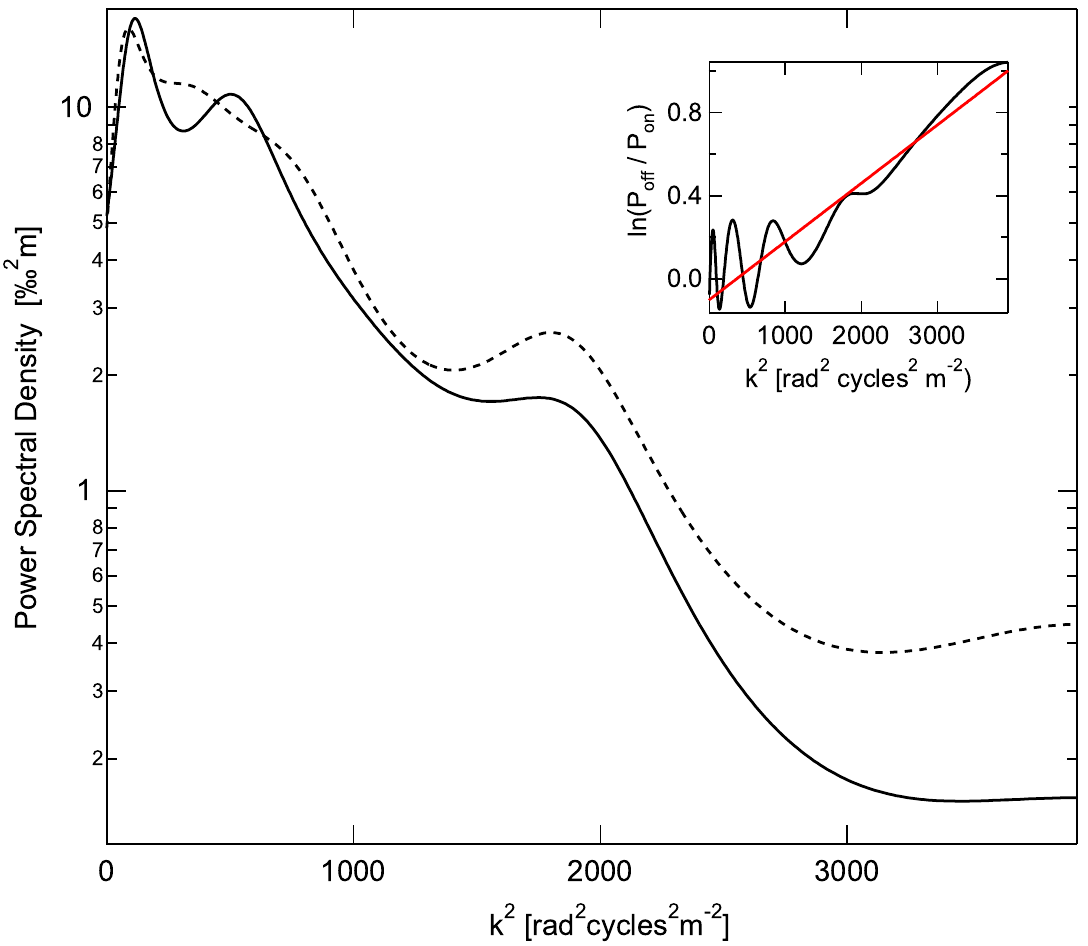}
\caption{Calculation of the diffusion length for the transfer function of the CFA system.
 The dashed and solid lines represent, respectively the power spectral density of the offline discrete data and the CFA data averaged on a~5\,cm resolution.}
\label{delsigmasq}
\end{figure*}

To overcome this problem we will present here an alternative way,
based on the comparison of the spectral properties of the on-line CFA
data and the off-line discrete data in 5\,cm sampling resolution,
presented in Sect.~3.2. In this approach the diffusion length of the
total smothing filter for the off-line discrete analysis will be:
\begin{equation}
\label{variance_off}
\sigma_{\rm off}^2 = \sigma_{\rm firn}^2 + \sigma_{\rm 5\,cm}^2 {}
\end{equation}
where $\sigma_{\rm 5\,cm}^2$ is the diffusion length of the smoothing
imposed by the sample cutting scheme at a~5\,cm resolution.  If one
averages the on-line CFA data at a~5\,cm resolution by means of
a~running mean filter, the diffusion length of the total smoothing
filter for the on-line CFA measurements averaged on a~5\,cm resolution
will be:
\begin{equation}
\label{variance_on}
\sigma_{\rm on}^2 =     \sigma_{\rm firn}^2 + \sigma_{\rm 5\,cm}^2 + \sigma_{\rm cfa}^2 {}
\end{equation}
From Eqs.~(\ref{variance_off}) and~(\ref{variance_on}) we get:
\begin{equation}
\label{delsigma}
\sigma_{\rm cfa}^2 = \sigma_{\rm on}^2 - \sigma_{\rm off}^2 {}
\end{equation}
As a~result, the term $\sigma_{\rm on}^2 - \sigma_{\rm off}^2$ is
directly related to the diffusion length of the smoothing filter of
the whole CFA-water isotope system including the melting and
debubbling sections. Based on Eq.~(\ref{attenuation}), the power
spectral density of the signals will be:

\begin{equation}
\label{power}
P = P_{0} {e}^{-k^2\sigma ^2} {}
\end{equation}
where $\sigma^2$ refers in this case to $\sigma_{\rm on}^2$ or
$\sigma_{\rm off}^2$.  Combining the power spectral densities of the
on-line and off-line time series we finally get:
\begin{equation}
\label{lnpower_plot}
\ln
\left( \frac{P_{\rm off}}{P_{\rm on}} \right) = \ln \left( \frac{P_{\rm 0off}}{P_{\rm 0on}} \right) + \sigma_{\rm cfa}^2 k^2  {}
\end{equation}
Hence, the logarithm of the ratio $P_{\rm off}/P_{\rm on}$ is linearly
related to $k^2$ with a~slope equal to $\sigma_{\rm cfa}^2$.  In
Fig.~\ref{delsigmasq} we perform this analysis for {\delD} and by
applying a~linear fit we calculate the $\sigma_{\rm cfa}[D]$ to be
equal to $16.4 \pm 2.4$\,mm. In a~similar manner
$\sigma_{\rm cfa}[O18]$ is found to be equal to $16.8 \pm 2.3$\,mm.

The higher value calculated with the spectral method points to the
additional diffusion of the sample at the melter and debubbler system
that could not be considered in the analysis based on the step
response. The impulse response of the system based on the updated
value of $\sigma_{\rm cfa}^2$ is presented in Fig.~\ref{Fig6}.

\subsection{Optimal filtering}

In the ideal case of a~noise-free measured signal
${\delta^{\prime}_{\rm m}}(z)$ and provided that the transfer function
$\hat{\mathcal{G}}_{\rm cfa}$ is known, one can reconstruct the
initial isotopic signal $\delta_{\rm i}(z)$ from Eq.~(\ref{Eq.2}) as:
\begin{equation}
\delta_{\rm i}(z) = \frac{1}{2\pi}\int_{-\infty}^{\infty}\frac{\hat{\delta}^{\prime}_{\rm m}
 (k )} {\hat{\mathcal{G}}(k )} e^{ik z}dk {}
\end{equation}
where the integral operation denotes the inverse Fourier transform and
$k = \frac{2\pi}{\lambda}$ with $\lambda$ being the wavelength of
the isotopic signals. In the presence of measurement noise $\eta(z)$,
this approach will fail due to excess amplification of the high
frequency noise channels in the spectrum of the signal.

Hereby we use the Wiener approach in deconvoluting the acquired
isotopic signals for the diffusion that takes place during the
measurement.  Considering a~measured isotopic signal
\begin{equation}
\delta_{\rm m}(z) = {\delta^{\prime}_{\rm m}}(z) + \eta(z) {}
\end{equation}
an optimal filter $\varphi(z)$ can be constructed that when used at
the deconvolution step, it results in an estimate of the initial
isotopic signal described as:
\begin{equation}
\tilde{\delta}_i(z) =\frac{1}{2\pi}\int_{-\infty}^{\infty} \frac{\hat{\delta}_{\rm m}(k
   )}{\hat{\mathcal{G}}(k )} \hat{\varphi}(k ) e^{i k z}dk {}
\end{equation}
Assuming that ${\delta^{\prime}_{\rm m}}(z)$ and $\eta(z)$ are uncorrelated
signals, the optimal filter is given by:
\begin{equation}
\hat{\varphi}(k ) = \frac{|\hat{\delta}_{\rm m}^{\prime} (k )|^2} {|\hat{\delta}_{\rm m}^{\prime} (k )|^2 + |\hat{\eta}(k )|^2} {}
\end{equation}
\citep{Wiener1949}; where $|\hat{\delta}_{\rm m}^{\prime} (k )|^2$ and
$|\hat{\eta}(k )|^2$ are the power spectral densities of the
signals $\delta_{\rm m}^{\prime} (z)$ and $\eta(z)$.

In the same fashion as in the previous section we assume that the
spectrum of the noise free measured signal $|\hat{\delta}_{\rm m}^{\prime}
(k )|^2$, is described by Eq.~(\ref{power}) where $\sigma^2 =
\sigma_{\rm tot}^2$.  Regarding the noise, we assume red noise
described by an AR1 process.  The spectrum of the noise signal will
then be described by \citep{Kay1981}:
\begin{equation}
|\hat{\eta}(k )|^2 = \frac{\sigma_{\eta}^2 \Delta z}
 {\left| 1+a_1 \exp{\left( -i k  \Delta z \right) } \right|^2} {}
\end{equation}
where $\sigma_{\eta}^2$ is the variance of the noise and $a_1$ is the
coefficient of the AR1 process.  We vary the parameters
\smash{$\sigma_{\rm tot}^2$}, \smash{$P_0$}, \smash{$\sigma_{\eta}^2$}
and \smash{$a_1$} so that the sum \smash{$|\hat{\delta}_{\rm m}(k
)|^2 = |\hat{\delta}_{\rm m}^{\prime}(k)|^2 + |\hat{\eta}(k
)|^2$} fits the spectrum of the measured signal.  The set of
parameters that results in the optimum fit is used to calculate the
optimal filter.

\begin{figure*}
\vspace*{2mm}
\center
\includegraphics[width=120mm]{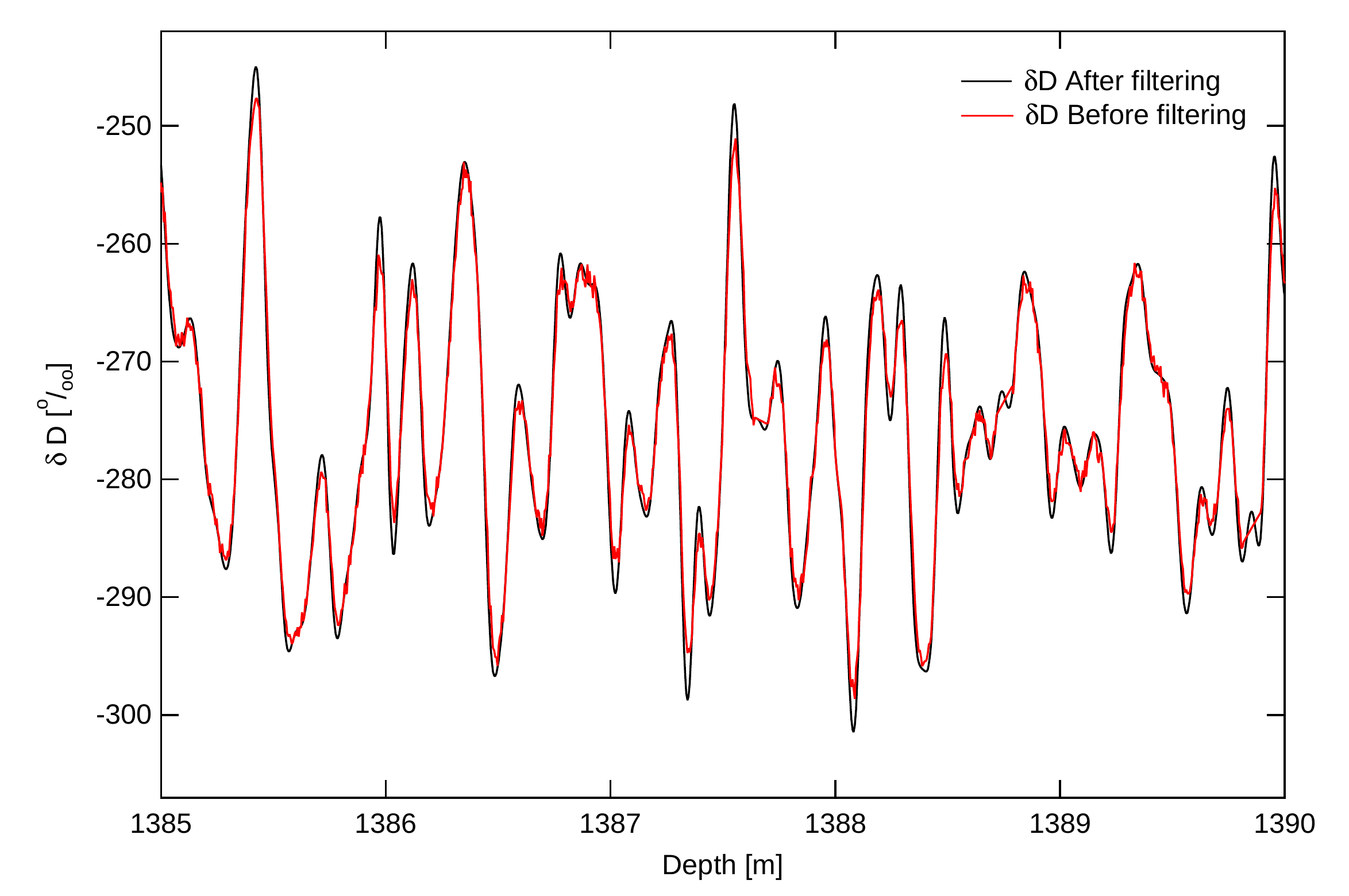}
\caption{{\delD} signal before and after optimal filtering.}\vspace{-4mm}
\label{Fig10}
\end{figure*}

The constructed filters together with the transfer functions that were
calculated based on the two different techniques outlined in
Sect.~\ref{section_resolution} are illustrated in Fig.~\ref{Fig8}.
One can observe how the restoration filters work by amplifying cycles
with wavelengths as low as 7\,mm.  Beyond that point, the shape of the
optimal filter attenuates cycles with higher frequency, which lie in
the area of noise.  An example of deconvoluted {\delD} data section is
given in Fig.~\ref{Fig10}.  It can be seen that the effect of the
optimal filtering results in both the amplification of the signals
that are damped due to the instrumental diffusion, as well as in the
filtering of the measurement noise.

\subsection{Information on deuterium excess}

\begin{figure*}
\vspace*{2mm}
\center
\includegraphics[width=150mm]{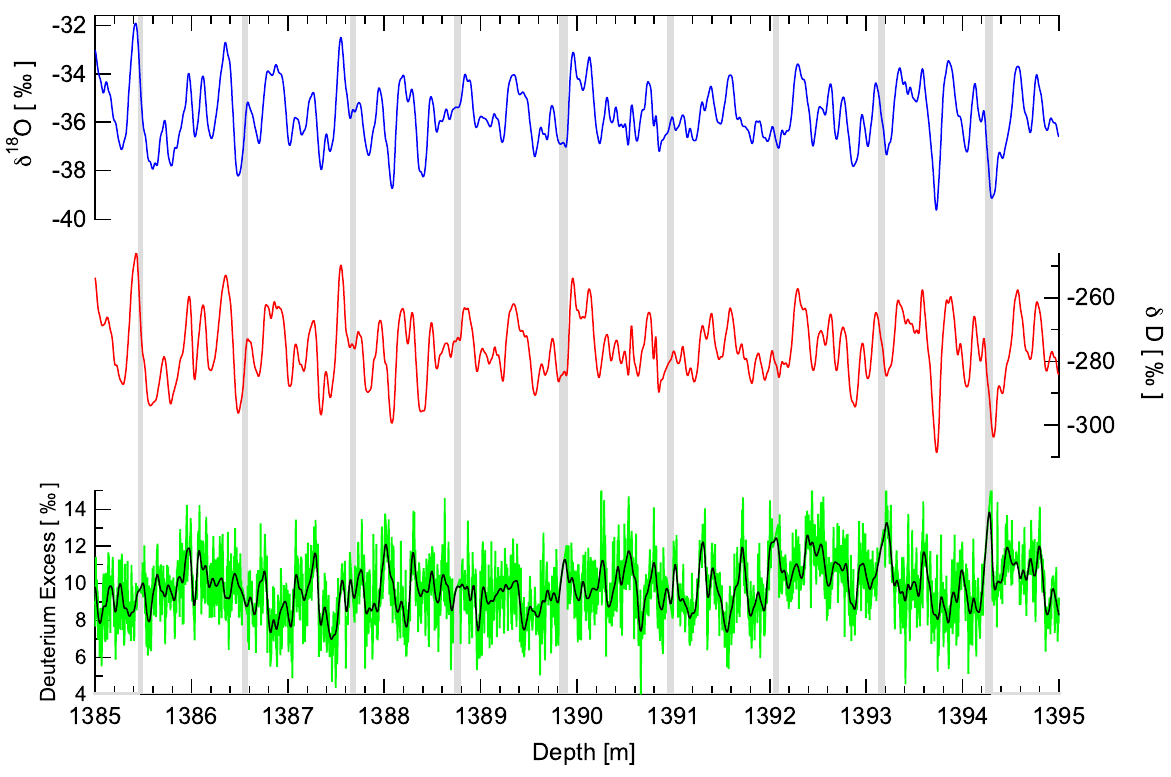}
\caption{{\delOx}, {\delD} and {\Dxs} signals after the optimal filtering. For the {\Dxs} we
  present the signal before (light green) and after (black) the filtering. Gray bars indicate
  the position and width of sections with missing/removed data.}\vspace{-4mm}
\label{Fig11}
\end{figure*}

Combining {\delOx} and {\delD} gives the deuterium excess as
{\Dxs}\,$=$\,{\delD}\,$-$\,8{\delOx} \citep{CRAIG1963, Mook2000}.  The noise
level of the {\Dxs} signal can be calculated by the estimated noise
levels of {\delOx} and {\delD} as:
\begin{equation}
{\sigma}_{\mathrm{D_{xs}}} = \sqrt{\sigma_{\rm D}^2 + 64\cdot {\sigma}_{18}^2} = 0.48 \,\,\,
\mbox{\permil}
\end{equation}
 As seen in Fig.~\ref{Fig11}, the {\Dxs} signal presents a~low signal
to noise ratio.  In this case, the technique of optimal filtering can
effectively attenuate unwanted high frequency noise components, thus
reveiling a~\qut{clean} {\Dxs} signal.

The latter offers the possibility for the study of abrubt transitions
as they have previously been investigated in {\delOx}, {\delD} and
{\Dxs} time series from discrete high resolution samples
\citep{Steffensen2008}.  The on-line fashion in which these
measurements are performed has the potential to yield not only higher
temporal resolution but also better statistics for those climatic
transitions.

\section{Conclusions}[Summary and conclusions]

We have succesfully demonstrated the possibility for on-line water
isotopic analysis on a~continuously melted ice core sample. We used an
infrared laser spectrometer in a~cavity ring down configuration in
combination with a~continuous flow melter system. A~custom made
continuous stream flash evaporator served as the sample preparation
unit, interfacing the laser spectrometer to the melter system.

Local water standards have been used in order to calibrate the
measurements to the VSMOW scale.  Additionally, dependencies related
to the sample size in the optical cavity have been accounted for. The
melting procedure is recorded by an optical encoder that provides the
necessary information for assigning a~depth scale to the isotope
measurements. We verified the validity of the applied calibrations and
the calculated depth scale by comparing the CFA measurements with
measurements performed on discrete samples in 5\,cm resolution.

By means of spectral methods we provide an estimate of the noise level
of the measurements.  The \mbox{combined} uncertainty of the measurement is
estimated at $\approx$0.06, 0.2, and 0.5\,{\permil} for {\delOx},
{\delD} and {\Dxs}, respectively.  This performance is comparable to,
or better than the performance typically achieved with conventional
IRMS systems in a~discrete mode.

Based on the isotopic step at the beginning of each CFA run, the
impulse response, as well as the transfer function of the system can
be estimated.  We show how this method does not take into account the
whole CFA system, thus underestimating the sample diffusion that takes
place from the melter until the optical cavity of the
spectrometer. We proposed a~different method that considers the power
spectrum of the CFA data in combination with the spectrum of a~data
set over the same depth interval measured in a~discrete off-line
fashion. The use of the optimal filtering deconvolution
technique, provides a~way to deconvolute the measured isotopic
profiles for apparent sample dispersion effects.

The combination of infrared spectroscopy on gaseuous samples with
continuous flow melter systems provides new possibilities for ice core
science.The non destructive, continuous, and on-line technique offers the
possibility for analysis of multiple species on the same sample in high
resolution and precision and can potentially be performed in the ﬁeld.

\section*{Acknowledgements}
We would like to thank Dorthe Dahl Jensen for supporting our research.
Numerous drillers, core processors and general field assistants have
contributed to the NEEM ice core drilling project with weeks of
intensive field work.  Withought this collective effort, the
measurements we present here would not be possible.  Bruce Vaughn and
James White have contributed to this project with valuable comments
and ideas.  This project was partly funded by the Marie Curie Research
Training Network for Ice Sheet and Climate Evolution
(MRTN-CT-2006-036127).\\
\\
Edited by: P.~Werle

\bibliographystyle{harvard}

\newpage
\newpage

\end{document}